\documentclass[11pt,twoside]{article}
\usepackage{./asp2014}

\aspSuppressVolSlug
\resetcounters

\begin{document}

\title{Complex Organic Molecules in Hot Molecular Cores/Corinos: Physics and Chemistry}

\author{Maria T.\ Beltr\'an and V\'{\i}ctor M.\ Rivilla
\affil{INAF-Osservatorio Astrofisico di Arcetri, Largo E.\ Fermi 5,
I-50125 Firenze, Italy; \email{mbeltran@arcetri.astro.it, rivilla@arcetri.astro.it}}
}

\paperauthor{Maria T.\ Beltr\'an}{mbeltran@arcetri.astro.it}{0000-0003-3315-5626}{INAF}{Osservatorio Astrofisico di Arcetri}{Firenze}{}{50125}{Italy}
\paperauthor{V\'{\i}ctor M.\ Rivilla}{rivilla@arcetri.astro.it}{0000-0002-2887-5859}{INAF}{Osservatorio Astrofisico di Arcetri}{Firenze}{}{50125}{Italy}


\section{Introduction}


Hot molecular cores (HMCs), the cradles of massive stars, are the most chemically rich sources in the Galaxy  \citep[]{bisschop07, belloche13, rivilla17}. The typical masses of these cores (few hundreds of\,$M_\odot$) make them the most important reservoirs of complex organic molecules (COMs), including key species for prebiotic processes. This rich chemistry is thought to be the result of the evaporation of dust grain mantles by the strong radiation of the deeply embedded early-type star(s). Our own Sun may have been born in a high-mass star-forming region  \citep[]{adams10}, so our Earth may have inherited the primordial chemical composition of its parental hot core region, as suggested by recent studies of oxygen and sulfur chemistry in comets  \citep[]{taquet16, drozdovskaya18}. 

The immediate surroundings of high-mass protostars are not the only regions rich in COMs. In recent years, the environments of some low-mass protostars have also revealed a very complex chemistry, with the presence of large molecules with abundances similar to those found towards high-mass protostars. These low-mass cores have been called hot corinos \citep[]{bottinelli04}. Although the mechanism responsible for the emission of COMs would be the increase of temperature in the inner envelope, like for the high-mass hot cores, it is not clear what is producing such an enhancement: shocks associated with infall or with outflows \citep[]{ceccarelli96, charnley05}. The fact that hot corinos are associated with low-mass young stellar objects allows us to study directly the environment where solar-analogs are now forming. 

In this chapter, we discuss how the ngVLA can help us to study the emission of heavy COMs in both low- and high-mass star-forming regions. The emission of COMs is   important not only because it allows us to understand how chemistry may have developed to eventually form life in our Earth, but also because COMs are a powerful tool for studying the physical properties and kinematics of the dense regions very close to the central protostars. 

\begin{figure*}
\includegraphics[width=\textwidth]{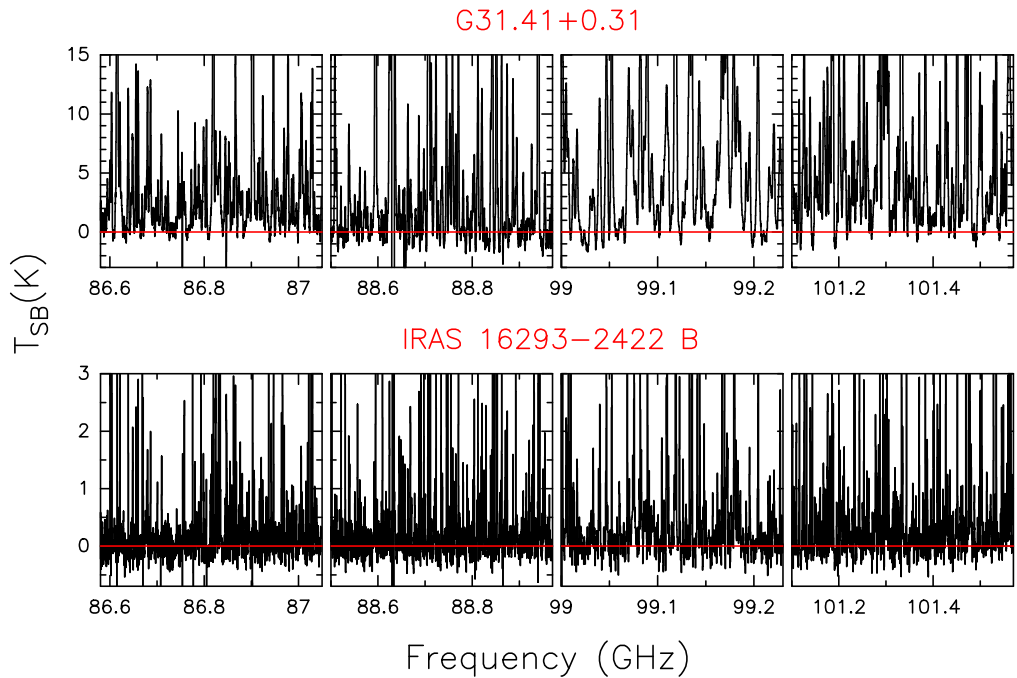}
 \caption{ALMA 3\,mm spectra at $\sim$1$''$ angular resolution of the hot molecular core G31.41+0.31 ({\it upper panels}) and the hot corino IRAS 16293-2422 B ({\it lower panels}) (V.\ M.\ Rivilla, private communication). Red lines indicate $T_{\rm SB}$=0\,K.  The 3-$\sigma$ noise level at each panel falls inside the red line. The (sub)millimeter spectra of chemically rich hot cores/corinos like these two examples are full of molecular lines, which complicates the detection of very complex molecules with weak emission due to line confusion limit and blending problems. }
\label{confusion}
\end{figure*}

\begin{figure*}
\begin{center}
\includegraphics[width=8cm]{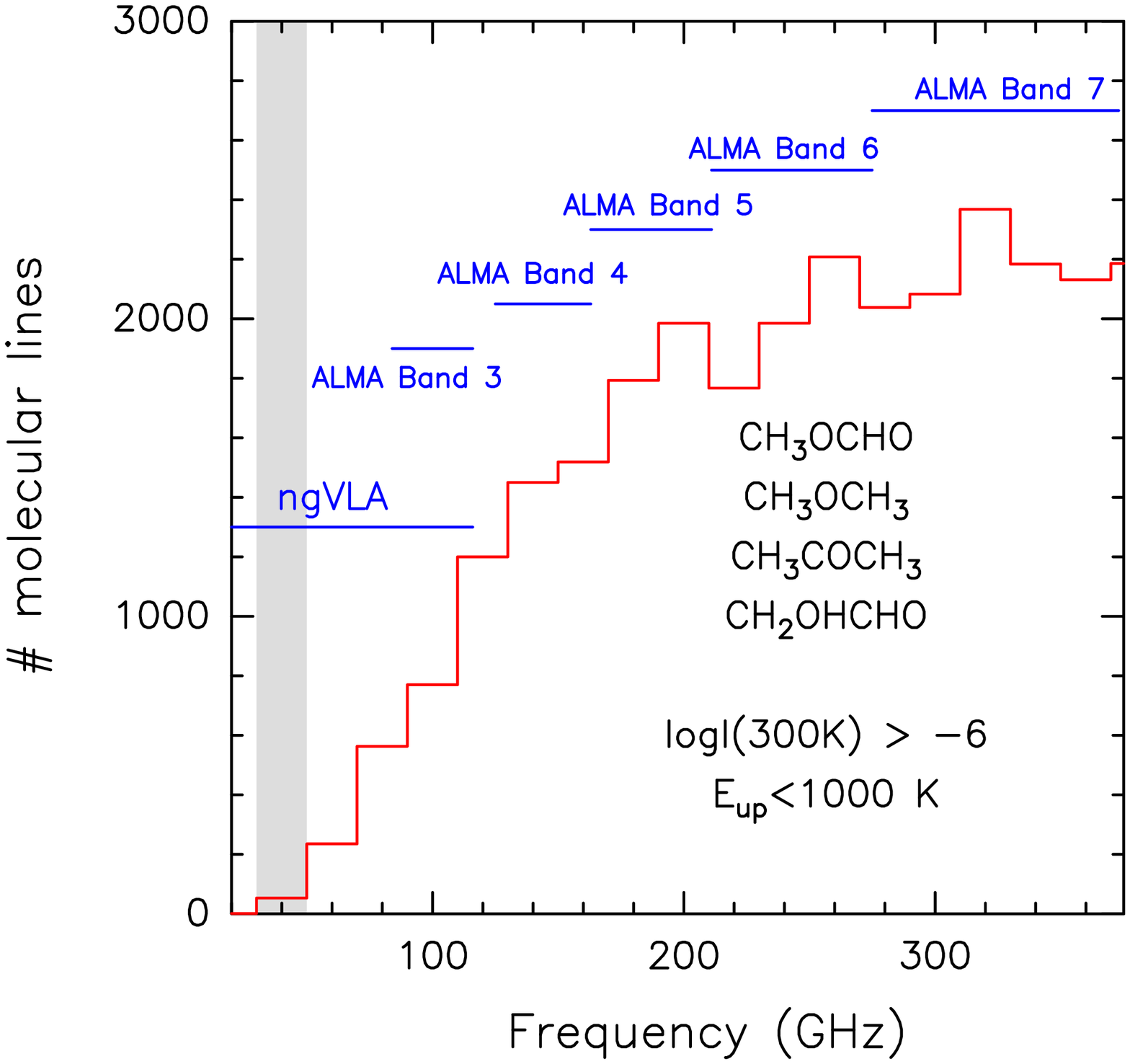}
 \caption{Distribution of the number of molecular lines as a function of frequency of four O-bearing COMs that are abundant in hot cores/corinos: methyl formate (CH$_3$OCHO), dimethyl ether (CH$_3$OCH$_3$), acetone (CH$_3$COCH$_3$), and glycolaldehyde (CH$_2$OHCHO). We have selected molecular transitions with $\log$\,I (300K)$> -6$ and E$_{\rm up}<1000$\,K from the JPL molecular catalogue. The frequency range shown in Figs.~\ref{fig-HNCHCN} and \ref{fig-MF}, 30--50 GHz, is indicated with a grey band. The spectral coverage of the  ngVLA and ALMA bands 3, 4, 5, 6 and 7 is shown.}
\label{comparison}
\end{center}
\end{figure*}

\section{COMs at centimeter wavelengths}

COMs in high- and low-mass star-forming regions can be observed at (sub)millimeter wavelengths but are severely affected by line blending, especially in turbulent regions where the line widths are broad.  In fact, many ALMA observations of the most chemically rich regions are essentially at the line confusion limit (Fig.~\ref{confusion}) -- that is,  no additional molecules can be detected that emit below that threshold regardless of the sensitivity (rms) of the spectrum.  Moving to the centimeter-wavelength window, the number of spectral lines in a selected spectral portion is significantly lower (see Fig.~\ref{comparison}) with respect to the millimeter window, and hence, the lines are less blended. Moreover, at millimeter wavelengths, the dust opacity is commonly very high, and may produce strong line absorption, preventing in some cases the detection of the line emission itself, while at centimeter wavelengths the dust emission is usually optically thin. 

The study of COMs at centimeter wavelengths is suited: 
\begin{itemize}

\item {\bf To detect heavy complex organic molecules in star forming-regions}. The PRIMOS survey, the GBT Legacy Survey of Prebiotic Molecules Toward Sgr B2N \citep[]{neill12, loomis13, zaleski13},  has detected in the last few years many COMs in this range in the massive HMC Sgr B2N, which is located in the Galactic Center.  Some examples are: propenal (CH$_2$CHCHO), propanal (CH$_3$CH$_2$CHO), simple aldehyde sugars like glycolaldehyde (CH$_2$OHCHO), the first keto ring molecule detected in a interstellar cloud, cyclopropenone (c$-$H$_2$C$_3$O), ketenimine (CH$_2$CNH), ethanimine (CH$_3$CHNH),  cyanomethanimine (HNCHCN) (see Fig.~\ref{fig-HNCHCN}), and ace\-ta\-mide (CH$_3$CONH$_2$), the largest interstellar molecule detected with a peptide bond, key for prebiotic chemistry.  Some of these species are, in fact, not suitable for observations at higher frequencies (millimeter and sub-millimeter regime). In general, the heavier the molecule, the more shifted to centimeter wavelengths its rotational spectrum is. In fact, these large molecules, with more than 8 atoms, have a considerable number of transitions at frequencies $< 50$\,GHz. 

\item  {\bf To study circumstellar disks surrounding the protostars}. Heavy COMs are in general less abundant than typical HMC tracers, such as methyl cyanide, and therefore, their emission is expected to be optically thinner. The combination of low dust opacity at centimeter wavelengths and of low line opacity of these large and low abundant  COMs, makes these species suitable to study the kinematics of high-mass star-forming regions -- in particular of possible accretion disks -- in a very pristine way. This probe is very important to understand better the formation of OB-type stars, for which the existence of true accretion disks is still under debate \citep[e.g.,][]{beltran16}. The kinematics of the gas close to the central protostar should tell us whether high-mass stars have Keplerian circumstellar disks, like those observed towards lower-mass counterparts (.g., L1527 IRS: Tobin et al.\ 2012; Ohashi et al.\ 2014; HD 163296: de Gregorio-Monsalvo et al.\ 2013), or whether they are surrounded by rotating structures undergoing solid-body rotation. 


\end{itemize}

\section{Measurements required}

The ngVLA will provide an unprecedented sensitivity and spatial resolution in the centimeter wavelength range that will allow us to detect new complex species and spatially resolve the emission of heavy COMs, e.g., those previously detected with the GBT (resolution $> 16''$). By targeting two of the most chemically rich star-forming regions, a high-mass hot molecular core \citep[G31.41+0.31: ][]{beltran09, rivilla17}  and a low-mass hot corino \citep[IRAS 16293-2422:][]{jorgensen16, martin17},  it will be possible to detect prebiotic species, such as HNCHCN (see Fig.~\ref{fig-HNCHCN}), for the first time outside the Galactic Center. In addition, the high-angular resolution provided by the ngVLA will allow us to map the distribution of heavy COMs from the envelope down to the accretion disks embedded in the hot core and hot corino, respectively, and characterize the region where the emission originates. It is important to understand whether the emission comes from the inner circumstellar disk where planets could form, or whether it originates in an outer region of the disk/envelope, where comets reside. 

In particular, the ngVLA capabilities would allow us to:

\begin{figure*}
\includegraphics[width=\textwidth]{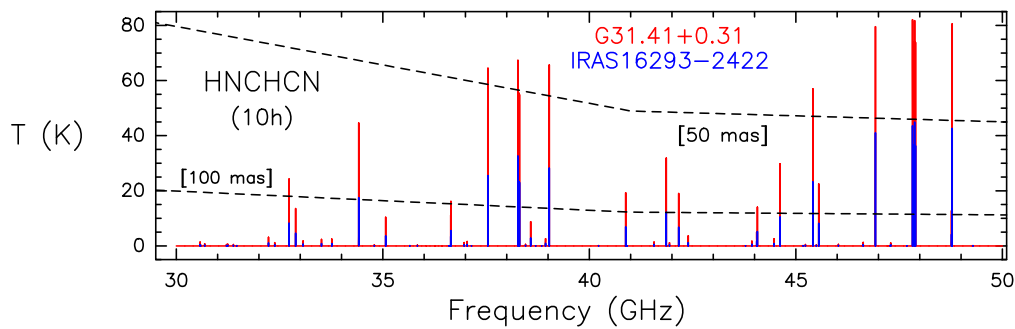}
 \caption{Synthetic spectra obtained with MADCUBA of cyanomethanimine, HNCHCN, for G31.41+0.31 hot molecular core (red) and IRAS16293-2422 B hot corino (blue). We have assumed an excitation temperature of 100 K, and column densities of 3$\times$10$^{17}$ cm$^{-2}$ and 3$\times$10$^{16}$ cm$^{-2}$ for G31.41+0.31 and IRAS\,16293$-$2422, respectively. The line widths considered are 5.0\,km s$^{-1}$ and 1.5\,km s$^{-1}$ for G31.41+0.31 and IRAS\,16293$-$2422, respectively, consistent with those found at milllimeter wavelengths \citep[]{rivilla17, jorgensen16}. We have overplotted with dashed lines the 3-$\sigma$ noise level for an integration time of 10 hr with two different angular resolutions, assuming a spectral resolution of 1 km s$^{-1}$, according to the specifications of the ngVLA.}
\label{fig-HNCHCN}
\end{figure*}

\begin{itemize}

\item {\bf Detect heavy (prebiotic) COMs, for the first time outside the Galactic Center}. 
First detections of new species in HMCs have been carried out mainly towards Sgr B2 \citep[see][]{belloche08, zaleski13}. This complex, located in the Central Molecular Zone, has been observed at different wavelengths and as a result, the first detections of heavy COMs in the Galaxy, such as ethylene glycol or ethyl formate (10 and 11 atoms, respectively), have been obtained there \citep[e.g.,][]{hollis00, hollis02}. The cores in Sgr B2, however, are not typical HMCs. The current star formation rate of Sgr B2 qualifies it as a mini-starburst region and its location close to the Galactic Center makes the physical conditions in that environment quite extreme, which could have significant consequences on the chemistry. Therefore, the chemistry in Sgr B2 might not be representative of that typical of HMCs in the disk of the Galaxy. In addition, the presence of several velocity components along the line of sight makes line identification more difficult. Therefore, to study the chemistry of COMs and the environments where they form, it is important to detect heavy COMs around more typical star-forming regions. The two star-forming cores G31.41+0.31 and IRAS\,16293$-$2422 are among the best candidates to detect new molecular species in high- and low-mass star-forming regions, respectively, outside the Galactic Center because of their chemical richness (see Fig.~\ref{confusion}).  They, thus, will serve as templates  for future observations of a range of sources with different stellar masses.  In particular, large molecules with $> 9$ atoms have been already detected in both regions \citep[][]{jorgensen16, rivilla17}. However, these two sources are not the only ones showing a chemically rich spectrum. HMCs are the cradles of OB stars, and it is therefore expected all high-mass (proto) stars to be associated with them \citep[][]{beltran10}. Regarding the low-mass regime, with the advent of ALMA, the number of known hot corinos is increasing 

To detect new species, high sensitivity is more important than high-angular resolution. In principle, it would not even be necessary to resolve the emission. In practice, however, it is important to resolve it,  at least slightly, to study the spatial distribution of the COMs and check whether the emission of different species is spatially correlated, i.\ e., coming from the same region.  

We have used the example of the prebiotic species cyanomethanimine (HNCHCN, see Fig.~\ref{fig-HNCHCN}), only detected so far by the GBT in Sgr B2N within the PRIMOS Project \citep[]{zaleski13}. To estimate the sensitivity (rms) required to detect emission from HNCHCN, we have modelled its emission assuming LTE conditions, an excitation of 100 K, typical of hot cores and hot corinos, and column densities of $3\times10^{17}$\,cm$^{-2}$ and $3\times10^{16}$\,cm$^{-2}$ for G31.41+0.31 and IRAS\,16293$-$2422, respectively (see caption of Fig.~\ref{fig-HNCHCN}). In this figure, we have plotted the 3-$\sigma$ noise level for an angular resolution of 50\,mas and 100\,mas and an integration time of 10 hr. As seen in the figure, at 50\,mas angular resolution, only emission from the high-mass hot molecular core G31.41+0.31 would be detectable. Since IRAS\,16293$-$2422 is located at only 120\,pc, however, the spatial resolution achieved even with 100\,mas ($\sim$12\,au), is already very high.
 
\begin{figure*}
\includegraphics[width=\textwidth]{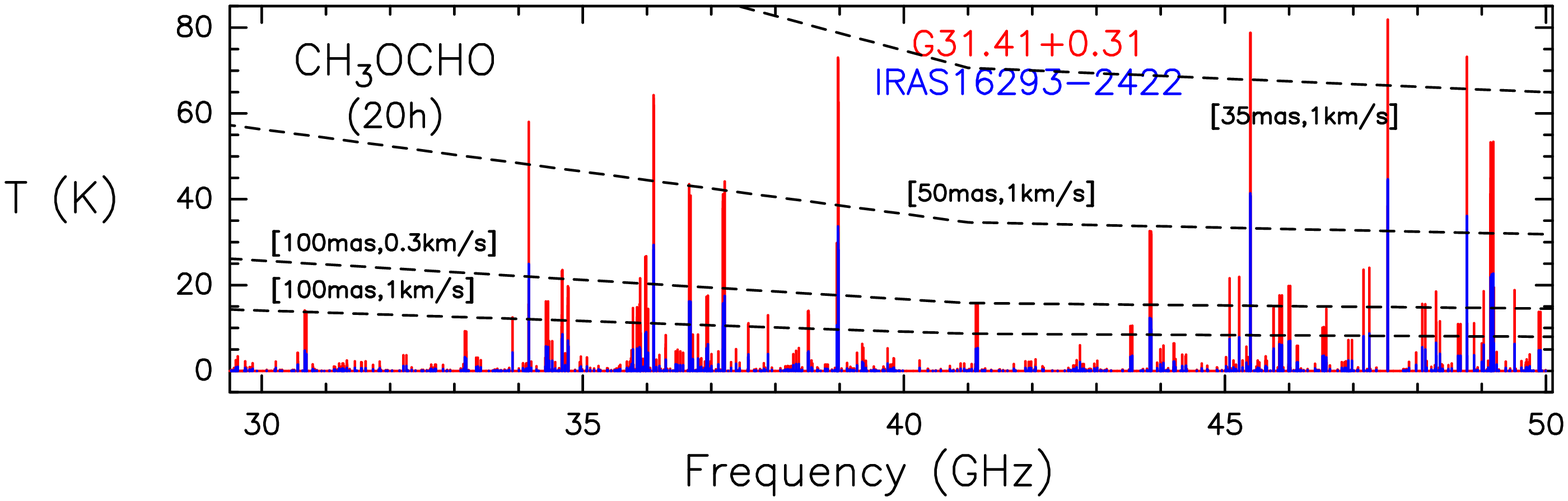}
 \caption{Synthetic spectra obtained with MADCUBA of methyl formate, CH$_3$OCHO, for G31.41+0.31 hot molecular core (red) and IRAS16293-2422 B hot corino (blue). We have assumed an excitation temperature of 100 K. The column densities for G31.41+0.31 and IRAS\,16293$-$2422 are 5$\times$10$^{18}$ cm$^{-2}$ and 5$\times$10$^{17}$ cm$^{-2}$, respectively, from Rivilla et al. (2017a) and Jorgensen et al. (2012). The linewidths considered are 5.0\,km s$^{-1}$ and 1.5\,km s$^{-1}$ for G31.41+0.31 and IRAS\,16293$-$2422, respectively. We have overplotted with dashed lines the 3-$\sigma$ noise level of an integration time of 20 hr with different combinations of angular and spectral resolutions, indicated with labels, according to the specifications of the ngVLA.}
    \label{fig-MF}
\end{figure*}

\item {\bf Study of the physical properties and kinematics of the circumstellar disks.} The distance to the solar-type protostar IRAS\,16293-2422 is 120\,pc. Therefore, with a spatial resolution of $\sim$5--10\,au, achieved with an angular resolution of 50--100\,mas, we should be able to study the kinematics and the chemistry of the inner disk region where planets could form. On the other hand,  the O-type (proto)star G31.41+0.31 is located much farther away, at 7.9\,kpc. By observing the core with an angular resolution of 35\,mas (a good compromise between angular resolution and sensitivity), the spatial scales achieved would be  
$\sim$280\,au, typical of disks around  intermediate-mass protostars (e.g., Beltr\'an \& de Wit 2016). Therefore, if a circumstellar disk is surrounding this massive star, the ngVLA should be able to detect it. Note that circumstellar disks have been detected up to late O-type stars \citep[e.g.,][]{johnston15}, but the luminosity of G31.41+0.31 ($3\times10^5$\,$L_\odot$) indicates that this HMC is hosting O6--O7 (proto)stars. Therefore, detecting such a disk would be crucial to confirm that stars of all masses form by disk-mediated accretion. 

To study the physics and kinematics of the disks, we have chosen the well-known high-density tracer methyl formate (CH$_3$OCHO, see Fig.~\ref{fig-MF}).  Methyl formate is a complex (8 atoms) molecule, which is a good thermometer and has been clearly detected in both regions with observations at much lower angular resolution \cite[]{jorgensen12, beltran18}. To estimate the required sensitivity, we have modelled the methyl formate emission assuming LTE conditions, an excitation of 100 K, typical of hot cores and hot corinos, and a column density of 5$\times$10$^{18}$ cm$^{-2}$ and 5$\times$10$^{17}$ cm$^{-2}$ for G31.41+0.31 and IRAS\,16293$-$2422, respectively (see caption of Fig.~\ref{fig-MF}), consistent with what found at millimeter wavelengths in the inner regions of hot cores/corinos \citep[]{jorgensen12, rivilla17}, and line widths of 5\,km\,s$^{-1}$ and 1.5\,km\,s$^{-1}$. The rms noise has been estimated for different angular and spectral resolutions. As seen in Fig.~\ref{fig-MF}, the methyl formate emission can be observed at high enough spatial resolution to study properly the kinematics of the circumstellar disks: up to $\sim$5\,au for IRAS\,16293$-$2422 and up to $\sim$280\,au for G31.41+0.31. 

\end{itemize}

\section{Synergy with other instruments}

Complex organic molecules can be observed from centimeter to millimeter wavelengths. As mentioned before,  to detect new species, high sensitivity is more important than high-angular resolution. Therefore, it is not surprising that some of the most recent detections have been done with a single-dish telescope, the GBT, within the PRIMOS Project \citep[][]{zaleski13}. However, although single-dish observations are a good starting point to detect new species, high-sensitivity interferometric observations are absolutely needed. On the other hand, because the emission of heavy COMs can be very weak and compact, it is not unusual that species previously not detected with a single dish, because of beam dilution, and easily detected with an interferometer. On the other hand, it is important to resolve the emission of COMs to study the spatial distribution of the COMs and check whether the emission of different species is spatially correlated, i. e., coming from the same region. This can help to constrain better the formation routes of different species and to distinguish between different chemical models (e.g., gas phase formation vs.\ grain surface). 

Regarding the millimeter regime, COMs have been detected with facilities such as IRAM NOEMA and ALMA, although as explain before, for chemically  rich sources, the millimeter observations can suffer from line confusion limit and line blending, which might difficult the identification of new species. In any case, millimeter and centimeter observations are complementary. By observing COMs at millimeter wavelengths with ALMA or NOEMA, we should be able to trace high-energy transitions of the same molecules and obtain a complete picture of the emission of such heavy species. This synergy will be very important to  constrain better the formation routes of heavy COMs.

\acknowledgements V.M.R. is funded by the European Union's Horizon 2020 research and innovation programme under the Marie Sk\l{}odowska-Curie grant agreement No 664931.  



\end{document}